\documentclass{aa}
\usepackage{graphicx}
\usepackage{txfonts}
\usepackage{natbib}
\bibpunct{(}{)}{;}{a}{}{,}

\newcommand{\be}{\begin{enumerate}}
\newcommand{\ee}{\end{enumerate}}

\def\mathstacksym#1#2#3#4#5{\def#1{\mathrel{\hbox to 0pt{\lower#5\hbox{#3}\hss} \raise #4\hbox{#2}}}}
\mathstacksym\gta{$>$}{$\sim$}{1.5pt}{3.5pt} 
\mathstacksym\lta{$<$}{$\sim$}{1.5pt}{3.5pt} 

\begin{document}
\title{The Origin of Massive O-type Field Stars}
\subtitle{Part II: Field O stars as runaways}
\titlerunning{Massive O-type Field Stars}
\author{        W.J. de Wit\inst{1}\thanks{Now at CNRS, Laboratoire d'Astrophysique de
                Grenoble} \and
                L. Testi\inst{1} \and
                F. Palla\inst{1} \and
                H. Zinnecker\inst{2}
                }
\offprints{W.J. de Wit, \email{dewit@obs.ujf-grenoble.fr}}
\institute{     INAF, Osservatorio Astrofisico di Arcetri, Largo E. Fermi 5,
                50125 Florence, Italy \and
                Astrophysikalisches Institut Potsdam, An der Sternwarte 16,
                14482 Potsdam, Germany}
\date{Received date; accepted date}
\abstract{In two papers we try to confirm that all Galactic high-mass
stars are formed in a cluster environment, by excluding that O-type
stars found in the Galactic field actually formed there.  In de Wit et
al. (2004) we presented deep K-band imaging of 5 arcmin fields centred
on 43 massive O-type field stars that revealed that the large majority
of these objects are single objects. In this contribution we explore
the possibility that the field O stars are dynamically ejected from
young clusters, by investigating their peculiar space velocity
distribution, their distance from the Galactic plane, and their
spatial vicinity to known young stellar clusters. We (re-)identify 22
field O-type stars as candidate runaway OB-stars.  The statistics show
that $4\pm2\%$ of all O-type stars with $\rm V<8^{m}$ can be
considered as formed outside a cluster environment. Most are
spectroscopically single objects, some are visual binaries.  The
derived percentage for O-type stars that form isolated in the field
based on our statistical analyses is in agreement with what is
expected from calculations adopting a universal cluster richness
distribution with power index of $\beta= 1.7$, assuming that the cluster
richness distribution is continuous down to the smallest clusters
containing one single star.
}
\maketitle

\section{Introduction}
The relatively brief existence of massive stars renders it likely that
the location where they form is where we observe them today. About
70\% of the massive O-type stars in the Galaxy is observed to be
associated with stellar clusters and/or OB-associations (Gies
1987\nocite{1987ApJS...64..545G}; Mason et
al. 1998\nocite{1998AJ....115..821M}; Ma{\'{\i}}z-Apell{\' a}niz et
al.  2004\nocite{2004ApJS..151..103M}). At least a third of the
remaining 30\% of the O-type stars are runaway OB-stars (Gies 1987),
and may therefore also have formed in a cluster, where it is known
that OB stars acquire high spatial velocities after dynamical
interactions or after supernova explosions in binary systems (Poveda et al.
1967\nocite{1967AJ.....72..824P}; van den Heuvel
1985\nocite{1985bems.symp..107V}; Gies \& Bolton
1986\nocite{1986ApJS...61..419G}; Clarke \& Pringle
1992\nocite{1992MNRAS.255..423C}; Hoogerwerf et al.
2001\nocite{2001A&A...365...49H}).  These statistical considerations
leave an absolute number of $\sim40$ O stars in the Solar
neighbourhood that are truly isolated, i.e. not known to be part of a
cluster/OB association nor to be runaway stars.
By retracing the formation history of these O-type field stars, 
we address in this paper the question whether a stellar cluster is a necessary
condition for the formation of a high-mass star.

A dense stellar cluster may provide favorable conditions for the formation of a
massive star through the process of coalescence of molecular cores in the very
early stages of star formation or even by merging stars in more evolved stages,
when the space density of these colliding objects exceeds some threshold value
(Bonnell, Bate \& Zinnecker 1998\nocite{1998MNRAS.298...93B}; Stahler, Palla \&
Ho 2000\nocite{2000prpl.conf..327S}; Bonnell \& Bate
2002\nocite{2002MNRAS.336..659B}; Bally \& Zinnecker 2005\nocite{zinn2005}). This suggests a physical connection between
the formation of a stellar cluster and a massive object, which is partly
supported by observations of the less massive, pre-main sequence Herbig AeBe
(HAeBe) stars. The gradual onset of clustering near these stars as a function of
spectral type seems to indicate that as the mass of the most massive object
increases, the richness of the associated cluster of lower mass stars also
increases (Testi, Palla, \& Natta 1999\nocite{1999A&A...342..515T}). However, as
pointed out by Bonnell \& Clarke (1999\nocite{1999MNRAS.309..461B}), this trend
does not imply that a cluster is a necessity for the formation of a high-mass
star. In general, the observed increase in richness of clusters with the mass
of the most massive HAeBe member was shown by these authors to be compatible
with random drawing of stellar clusters that are distributed in membership
number according to a certain power law, and subsequently populated with stars 
following a universal stellar IMF.

Statistical IMF arguments allow for a finite probability of forming a massive
star without the required cluster and indirectly provides support for the
formation of a massive star set by conditions other than a stellar cluster, e.g.
a sufficiently massive, non-fragmenting dense molecular core and an accretion
disk (e.g. Yorke \& Sonnhalter 2002\nocite{2002ApJ...569..846Y}; Li et al.
2003\nocite{2003ApJ...592..975L}). Accretion disks near young massive stars have
recently been suggested by mm observations (Beltran et al. 2004\nocite{0312495})
and the near infrared (Chini et al. 2004\nocite{2004Natur.429..155L}). 
Modeling of the near infrared CO bandhead emission may also indicate that a
fraction of young massive stars is surrounded by rotating Keplerian disks (Bik
\& Thi 2004\nocite{2004A&A...427L..13B}; Bik, Kaper \& Waters
2005\nocite{Bik2005}). Isolated formation of single massive stars could be
occurring in external Galaxies (Large Magellanic Cloud and M\,51, Massey et
al. 1995\nocite{1995ApJ...438..188M}; Lamers et al.
2002\nocite{2002ApJ...566..818L}).

The field O-type star population provides the opportunity to test the ``null
hypothesis'' that all massive stars form in stellar clusters. The validity of
this hypothesis was addressed already in 1957 by Roberts
(1957\nocite{1957PASP...69...59R}). Within this context, we try to elucidate
here the formation history of the field O stars. Their properties in terms of
location, radial velocity and binarity were already presented by Gies (1987,
hereafter G87). He finds that field O stars have characteristics intermediate
between O stars in clusters/associations and the runaway O stars, and suggested
that a substantial fraction of them could belong in fact to the latter
category. However, the possibility that some are actually formed at their
present (isolated) location in the Galactic field is still open. In this second
contribution of our series, we explore the possibility of a runaway nature that
would explain their location in the Galactic field.

The paper is organized in the following way. In Sect.\,\ref{prev} we give a
summary of the observational results obtained in de Wit et al. (2004, hereafter
Paper\,I)\nocite{2004A&A...425..937D}, in which we searched for the presence of subparsec scale 
stellar clusters near field O stars.
In Sect.\,\ref{deriv} we explore the dynamical ejection scenario for
the origin of field O stars as a group by determining space velocities
using Hipparcos proper motions, the distance to the Galactic plane,
and the presence of nearby young clusters. We discuss our statistical results in 
Sect.\,\ref{disc} and compare these with theoretical expectations. The main
conclusions are summarized in Sect.\,\ref{concl}.

\section{Main results from Paper\,I}
\label{prev}
The observational objective of Paper\,I (de Wit et
al. 2004\nocite{2004A&A...425..937D}) was to determine by deep imaging
the presence of small scale stellar clusters near the sample of field
O-type stars (defined in Mason et al. 1998), as is found to be the case 
near the HAeBe stars (Testi et al. 1999). Given the low detection rate, we
dismissed the possibility that the formation of a field O-type star proceeds in
small stellar clusters, under our initial assumption that all the field O stars
in the sample were actually formed in the Galactic field.

\subsection{The absence of clusters near field O stars}
We selected all 43 O-type stars characterized as field objects from the
interferometric multiplicity study of $\sim$200 O stars by Mason et al. (1998,
hereafter M98). We searched for hitherto unknown stellar clusters centred on the
target stars using stellar density maps. High resolution density maps were
constructed from deep K-band images taken with NTT/SOFI and TNG/NICS, probing
linear scales of $\rm \sim0.25\,pc$. Lower resolution maps were constructed from 2MASS
K-band covering tens of parsecs with a linear resolution of $\rm \sim1.0\,pc$. A
3$\sigma$ deviation from the average stellar density was considered to be a
cluster provided that it was centred on the target star. The maps are 
presented in Paper\,I along with the K-band images. In 5
cases we detect a clear stellar density enhancement near the field O star, four
of which were previously thought to be visually single objects. 

Paper\,I briefly describes each field O-type star, with the emphasis on their
spectroscopic and visual multiplicity status, and the presence of star formation
indicators in the field like IRAS sources, H\,{\sc ii} regions or dark
clouds. Specific attention is given to runaway stars. Given the objective of the
project to determine whether isolated massive stars form in an isolated way,
a clear division is applied between field stars and runaway objects. In
fact the M98 characterization of a field object depends on the
strict definition of a runaway star (requirements regarding the radial velocity
and the distance from the Galactic plane) and not on the star's (possible) birth-site
as is appropriate for this study.  For example, according to M98 the well known
runaway objects $\zeta$~Pup and HD\,75222 are considered field stars owing to
their rather small radial velocity component. Within the framework of our study,
we prefer to classify runaway stars as cluster members, and reserve the term ``field 
star'' for the objects that do not find an origin in a cluster.

\begin{table}[t]
 {
  \begin{center}
\caption[]{Field O stars in newly detected clusters. The asterisk indicates that
the star is found in a region with star formation signposts (see
Sect.\,\ref{youth}). In Col.\,2 we give the visual multiplicity (V.M.) of the stars,
adopted from M98 and G87; VMS = Visual Multiple Star. In Col.\,3 we use the
following abbreviations for the spectroscopic multiplicity (S.M.): C = constant radial
velocity, SB1 = single-lined spectroscopic binary, SB2 = double-lined
spectroscopic binary. Addition of an "O" and/or "E" indicates that the orbit is 
known and/or an eclipsing system. A colon indicates uncertainty.}

  \begin{tabular}{lccccc}
\hline
\hline
Name & V.M. & S.M. & $\rm I_C$ & Radius  & $\log(\rho_{\rm I_{C}})$\\
     &            &             &           &    (pc) & ($\rm stars\,pc^{-3}$) \\
\hline
HD\,52266$^{*}$   & single & SB1? & $4\pm2$  & 0.10 & $2.0\pm1.0$ \\
HD\,52533$^{*}$   & VMS    & SB1O & $15\pm5$ & 0.30 & $2.5\pm0.9$  \\
HD\,57682         & single & C    & $4\pm5$  &  -   & - \\
HD\,153426        & single & SB2: & $5\pm4$  &  -   & -\\
HD\,195592$^{*}$  & single & SB1? & $18\pm3$ & 0.25 & $2.6\pm0.8$ \\
\hline
 \end{tabular}
\label{sfreg}
\end{center}
}
\end{table}

If we would subdivide the 43 ``field O stars'' according to their visual
multiplicity (based on G87 and M98), we get the following picture: 27 single
objects, 5 optical binaries, 6 visual binaries, 2 visual multiple objects, and 3
runaway stars ($\zeta$~Pup, HD\,75222, and the runaway X-ray binary
HD\,153919; Ankay et al. 2001\nocite{2001A&A...370..170A}). It is noteworthy that the vast majority
are visually single objects.  Since the three runaways are obviously not formed
where they are currently located presently, we will not discuss their properties
further, although we will keep them in the statistics.

The main result of Paper\,I is that the majority of the massive O stars are
found not to be associated with clusters on scales of a few tenths to a few parsec. In
fact, the stellar density maps of Paper\,I show that at least $\sim 85\%$ of the high
mass field stars are isolated objects. In only 5 cases we detected a cluster.

\subsection{The 5 field O stars in newly detected clusters}
\label{new}
Using deep infrared imaging, we detected small scale clusters
associated with 5 field O stars (see Table\,\ref{sfreg}). We quantify
the stellar richness of these clusters using the $\rm I_C$ indicator
introduced by Testi et al. (1997\nocite{1997A&A...320..159T}) in their
study of clusters near HAeBe stars.  The $\rm I_C$
parameter estimates the stellar {\it richness} of the cluster centered
on the O star by counting the number of stars corrected for the
back/foreground field contamination within the determined, assumed
circular, cluster radius. 
The properties of the new clusters are given in Table\,\ref{sfreg}.

The table reflects what is clear from the stellar density maps: the stars
HD\,52533 and HD\,195592 are located in the richest cluster among the field O
stars. HD\,52533 was already known to be a visual multiple system, consisting
of at least 4 components (M98), whereas HD\,195592 has been catalogued as a
visually single object (possibly a single line spectroscopic binary).
The cluster detection for the other three stars is marginal as is 
reflected by their low $\rm I_C$ values. 

In Col.\,5 of Table\,\ref{sfreg}, we list the cluster radii for the cases with
an approximately spherical density enhancement centered around the field O
star. The average value of $\sim 0.2$\,pc is remarkably similar to that of the
clusters around HAeBe stars (Testi et al. 1999). We have used this average
value for the radius to convert the $\rm I_C$ value to stellar spatial densities.
We conclude that both the richness and extent of especially the clusters found
near HD\,52533 and HD\,195592 show that what is observed among the highest mass
HAeBe stars.  In Appendix\,\ref{appb} we describe the details concerning the
clusters found near these two stars.

\section{The field O stars as candidate runaway stars}
\label{deriv}
\begin{figure}[t]
\includegraphics[height=8cm,width=8cm]{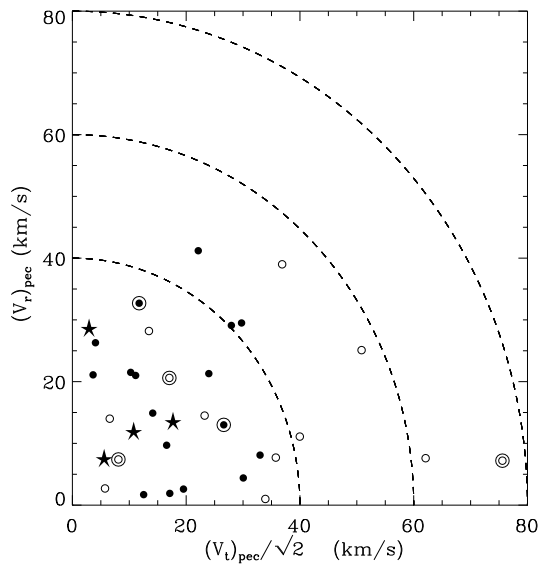}
  \caption[]{The space velocity of field O stars. Note that not all 
  field O stars under study have Hipparcos proper motions. The peculiar tangential
  velocities of the field O stars are derived from Hipparcos proper motions.
  The uncertainty has been calculated by propagating a 30\% uncertainty in
  the distance. Symbols are as follows: filled asterisks indicate absolute
  uncertainty less than $\rm 10$ km\,s$^{-1}$, filled circles an uncertainty
  between 10-20 km\,s$^{-1}$, and empty circles $\rm >$20 km\,s$^{-1}$.  Some
  field O stars have space velocities negligibly different from the bulk of
  the Galactic field. Five stars with clusters discussed in Paper\,I are
  encircled.}
\label{vpec2}
\end{figure}

Our null hypothesis that massive stars are all formed in clusters requires that
the absence of clusters near the Galactic massive field population is
interpreted in terms of a dynamical ejection scenario, similar to (part of) the
runaway OB stars (e.g. Hoogerwerf et
al. 2001\nocite{2001A&A...365...49H}). In this section we test this possibility
by examining the spatial
velocities of our sample using Hipparcos (ESA 1997)\nocite{1997yCat.1239....0E} proper motion measurements and 
published radial velocities, their distances above the Galactic plane, 
and the existence in the literature of very young stellar clusters with 
sufficient properties to be capable of producing high-mass runaway stars.

\subsection{Space velocities}
\label{space}
Knowledge of
the spatial velocities of the field O stars can be used to exclude a runaway
history on kinematic grounds. Traditionally, the minimum peculiar velocity for
classifying a runaway star is $\rm 40\,kms^{-1}$ (Blaauw
1961\nocite{1961BAN....15..265B}). The radial component of the peculiar space
velocity of the O field stars is discussed in G87 who found that the
average of $\rm <(v_{r})_{p}>=6.5$ km\,s$^{-1}$ is generally lower than
for the runaways, but higher than that of O stars in clusters/OB
associations. Below, we derive the motion in the plane of the sky from
Hipparcos measurements of our targets.

Hipparcos proper motions are available for 34 stars of our sample. We derive the
corresponding tangential peculiar velocities using the distance estimates taken
from G87 and M98, as listed in Paper\,I.  Following Moffat et al.
(1998\nocite{1998A&A...331..949M}), we adopt a flat rotation
curve with a Solar galactocentric distance of 8.5\,kpc and a circular Galactic
rotation velocity of $\rm 220\,km\,s^{-1}$. The adopted Galactic rotation model
should be adequate for galactocentric distance between 3 and 18\,kpc (Kerr \&
Lynden-Bell 1986\nocite{1986MNRAS.221.1023K}). A 30\% uncertainty in
distance is propagated in the error estimate of $\rm <(v_{t})>$.

\begin{table}[t]
 {
  \begin{center}
 \caption[]{Runaway O-type candidate stars derived from Hipparcos proper motions
  and distance from Galactic plane (Z). The abbreviation in Col.\,2 and 3 are
  the same as in Table\,1 and OPT = optical binary, VB = visual binary. The
  criterion for classification of the listed stars as candidate runaway stars is given in
  the last column.}
  \begin{tabular}{lccc}
\hline
\hline
Name & V.M. & S.M. & Remark\\
\hline
HD\,1337  & single & SB2OE & Z \\   
HD\,15137 & single & SB2? & $\rm v_{pec}$ + Z \\ 
HD\,36879 & single & C    & $\rm v_{pec}$\\
HD\,41161 & VB     & C     & Z \\   
HD\,57682$^1$ & single & C    & $\rm v_{pec}$\\
HD\,60848 & single & U     & Z \\   
HD\,89137 & single & SB1?  & Z \\   
HD\,91452 & single & C    & $\rm v_{pec}$ + Z\\ 
HD\,105627& single & C    & $\rm v_{pec}$\\
HD\,122879& single & C    & $\rm v_{pec}$\\
HD\,163758& single & C     & Z \\   
HD\,175754& single & C     & Z \\   
HD\,175876& OPT    & C     & Z \\   
HD\,188209& single & C     & Z \\   
HD\,201345& single & C    & $\rm v_{pec}$ + Z\\ 
\hline
\multicolumn{4}{l}{\tiny 1. The star is also found to have a cluster in
  Sect.\,\ref{prev}. See text.}
 \end{tabular}
\label{runw}
\end{center}
}
\end{table}

The resulting $\rm (v_{t})_{p}$ normalized to one component against
$\rm (v_{r})_{p}$ is presented in Fig.\,\ref{vpec2}. The values of
$\rm (v_{r})_{p}$ are taken from G87. The error is dominated by the
distance uncertainty.  Filled asterisks are stars with the most
accurate measurements and an absolute uncertainty less than $\rm
10$\,km\,s$^{-1}$; dots have an absolute uncertainty between 10 and
20\,km\,s$^{-1}$, and empty circles are for uncertainties in excess of
20\,km\,s$^{-1}$ (see the caption to Fig\,\ref{vpec2}). The five encircled symbols correspond to the stars
in Table\,1.  Note that the relative uncertainty is smallest for the
star with the largest absolute uncertainty, i.e. the suspected runaway
HD\,57682 (see de Zeeuw et al. 1999\nocite{1999AJ....117..354D}). Also
note that stars with low velocities have uncertainties larger than
their measurements.  Therefore, the lower velocity part of the diagram
is not very useful except that it makes clear that most field O stars
do not have high velocities.

Fig.\,\ref{vpec2} demonstrates that by applying a strict runaway limit
of $\rm 40\,kms^{-1}$ regardless of the error bars, the Hipparcos
proper motions allow the identification of 7 additional candidate
runaway OB stars, one of which with a {\it radial} velocity component
marginally larger than $\rm 40\,km\,s^{-1}$.  They are listed in
Table\,\ref{runw}.  While some of them have already been suggested as
runaways, they were not classified as such by G87 due to the more
stringent selection criteria applied; these stars were therefore
instead classified in the more general terms of an O-type field star.
Apart from the double line spectroscopic binary HD\,15137, the other
six stars are optical/spectroscopic singles (see
Table\,\ref{runw}). This would corroborate the hypothesis of a
dynamical origin, because a high velocity ejection of a multiple 
system is not likely.

Finally, we note that the four stars with clusters (encircled symbols)
do not have large spatial velocities. They occupy the region of 
the field population in Fig.\,\ref{vpec2}. A exception is HD\,57682, 
that has the largest proper motion of all. It therefore becomes probable 
that the cluster marginally detected is likely to be a statistical noise
fluctuation (see also Paper\,I).

\subsection{Distance from the Galactic plane}
\label{zdist}
\begin{figure}[t]
\includegraphics[height=8cm,width=8cm]{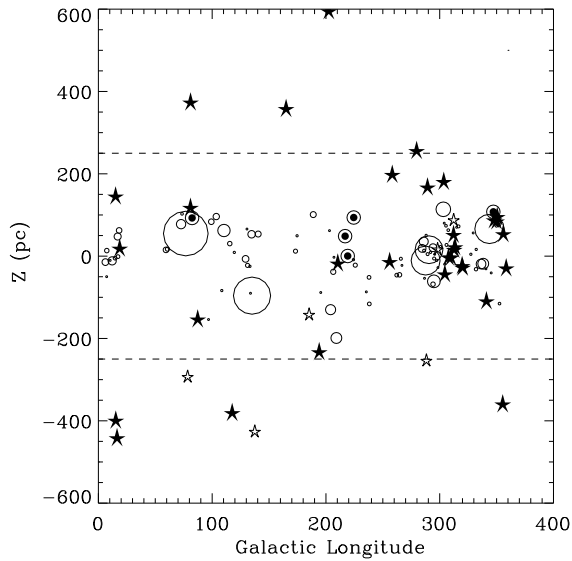}
  \caption[]{Height above the Galactic plane (Z) as function of the galactic
  longitude for O field stars. The empty asterisks are suspected runaway
  stars from Table\,\ref{runw} and stars associated with clusters of
  Table\,\ref{runw2}. The five filled and encircled symbols have indication of a 
  nearby cluster. All open circles indicate associations of OB stars within 3\,kpc
  of the Sun from Mel'Nik \& Efremov (1995\nocite{1995AstL...21...10M}). The
  size of each open circle is proportional to the estimated richness of the
  association.
}
\label{zdistr}
\end{figure}
Distance from the Galactic plane (Z) is a useful and indirect tool to infer
present (or past) high spatial velocities. The threshold distance for
identifying runaway OB star is generally taken to be 500\,pc (G87). This
translates into a perpendicular Z-component velocity of $\rm \sim30$ km
s$^{-1}$, given that high-mass star formation is confined to $\sim$200\,pc from 
the Galactic plane, as illustrated in Fig.\,\ref{zdistr}.
The figure shows the OB associations within 3 kpc radius of the Sun identified
by Mel'Nik \& Efremov (1995). The OB associations are represented by the circles 
whose size depends on the stellar richness. Star formation is assumed not to occur in the halo of Galaxy (e.g. Ramspeck 
et al. 2001). 

To exclude possible OB-runaway stars from our sample, 
we adopt a threshold value of $\rm Z=250\,pc$ as the
maximum allowed distance for the field O stars. This number is derived
from the observed distribution of OB association and the ``drift
distance'' of $\sim$65\,pc based on the O star main-sequence
lifetime. Field O stars have an average peculiar radial velocity of
$\rm 6.4$ km s$^{-1}$ (G87) and, moving ballistically, may therefore
wander about 65\,pc in an average lifetime of $10^{7}$ yr.

Any O field star with Z-value larger than 250\,pc is regarded as a
candidate OB-runaway star. This strict constraint leads to the
exclusion of 11 O-type stars, marked as asterisks in Fig.\,\ref{zdistr} that are
above and below the dashed threshold distance. Among these, three stars
(HD\,15137, HD\,91452, HD\,201345) are also found to have 
large spatial velocities (the open asterisks). 

As an example, we point out the remarkable star HD\,41161 at a distance
of $\rm Z=355\,pc$ and at $ l=165\deg$. At this distance from the Galactic plane 
this star is an OB-runaway candidate. This star shows evidence for a IR bow shock
(Noriega-Crespo et al. 1995), which would strongly confirm its runaway
status. However, HD\,41161 is listed as a visual binary with 10 arcsec
separation and a period of 243\,000 yr (M98). Whether hard binaries
may be able to dynamically eject a visual binary is very unlikely
(Leonard \& Duncan 1990\nocite{1990AJ.....99..608L}).  On the other
hand, wide binaries are readily destroyed by stellar encounters in
dense clusters. The nature of HD\,41161 remains therefore unclear.

\begin{table*}[t]
 {
  \begin{center}
 \caption[]{Field O stars possibly associated with young clusters. See Tables\,\ref{sfreg} and
 \ref{runw} for the meaning of the abbreviations used in Col.\,2 and 3. The
 projected distance is with respect to the distance of the field O
 star.}
  \begin{tabular}{lcccccc}
\hline
\hline
Name & V.M. & S.M.    & Proj. dist.& Cluster     & Age & Ref\\
     &            &                &   (pc)     &             & (Myr) & \\
\hline
HD\,117856&  VB     & SB2:  & 57 & St\,16         & 4  & 1\\
HD\,125206&  single & SB2:  & 45 & NGC\,5606     & 7   & 2\\
HD\,154368&  VB     & SBE   & 42 & Bo\,13        & 6.3 & 3\\
HD\,154643& single  & SB1:  & 44 & Bo\,13        & 6.3 & 3\\
HD\,158186& single  & SBE   & 30 & NGC\,6383     & 1.7 & 4\\
HD\,161853& single  & SB1:  & 55 & Col\,347      & 6.3 & 3\\
HD\,169515& single  & SB2OE & 65 & NGC\,6604     & 4   & 3\\
\hline
 \end{tabular}
\label{runw2}
\end{center}
\hspace {3.65cm}\vspace{-0.1cm} {\tiny 1: Turner 1985\nocite{1985ApJ...292..148T}; 2: Vazquez et
al. (1994)\nocite{1994A&AS..106..339V}; 3: Battinelli et
al. (1994)\nocite{1994A&AS..104..379B}; 4: Fitzgerald et
al. (1978)\nocite{1978MNRAS.182..607F}}
}
\end{table*}

\subsection{Low velocity ejection from young stellar clusters}
\label{lowv}
The observations presented in paper\,I show that less than 12\% of the Galactic field O
stars are found in small clusters. In the previous two subsections, we have
seen that 15 additional stars can be considered as possibly dynamically
ejected stars based on a high tangential peculiar velocity or a large
distance from the Galactic plane. Here, we consider the possibility that the
field O stars in the Galactic plane ``ran away'' from their birth clusters at
a more moderate velocity than the classical runaway stars. However lacking
precise proper motion data, we attempt to make basically a
zero-order retracing of the field O stars back to their possible location of origin.
We adopt the simple assumption that finding young clusters in the vicinity of an
isolated O-type star supports the conjecture that the star originates there. 
The tangential peculiar velocities derived in the previous subsection are
useful only for stars with high proper motions, while for the low tangential
peculiar velocities the relative error is generally too large to allow an
estimate of the direction of the stellar motion.

In Paper\,I we report for each O star the clusters with an age less
than 10\,Myr that are found within the projected ``drift distance'' of
65\,pc (see previous section). We require that the distance to the Sun
of the O star and cluster are similar within the uncertainty and adopt
a distance uncertainty for the O stars of 30\%. In 7 cases a young
cluster is found to exist within the set boundaries.  In
Table\,\ref{runw2} we list the basic information on each O-type field
star and associated cluster. In these cases we assume that the
probability to find such a cluster and an O field star in the
same region of the sky is small. Below we describe briefly the characteristics
of each cluster, as far as they are described in the literature:
\smallskip
\newline
\noindent{\bf Stock\,16:} A massive young cluster lying in the \ion{H}{ii} region 
\object{RCW\,75} containing a 09.5 V star \object{HD\,115071} (Penny et al. 2002\nocite{2002ApJ...575.1050P}) 
and for which the most luminous member is identified as an O7.5III((f))
star by Walborn (1973)\nocite{1973AJ.....78.1067W}. This particular cluster stands out due to a
large proportion of close binary systems among its present population
(Dokuchaev \& Ozernoy 1981\nocite{1981SvAL....7...52D}) indicating possibly a dynamical
past with frequent gravitational encounters between its
members.\newline
{\bf NGC\,5606:} The core of this cluster is reported by Vazquez et
al. (1994)\nocite{1994A&AS..106..339V} to be underpopulated in low-mass
stars and showing ``five bright stars forming a compact group''. 
Such a hierarchical situation is expected to exist in a cluster able to expel
massive stars (Clarke \& Pringle 1992).\newline
{\bf Bochum\,13:} No characteristics known. \newline
{\bf NGC\,6383:} In the centre of the young cluster a spectroscopic binary 
\object{HD\,159176} (O7 V + O7 V, P = 3.37 days; Stickland et al. 1993\nocite{1993Obs...113..204S})
is found. The cluster may belong to the \object{Sgr OB1} association.\newline
{\bf Col\,347:} No characteristics known. \newline
{\bf NGC\,6604:} The young cluster contains some massive components in the form
of an eclipsing double-line spectroscopic  O5-8V + O5-8V binary with a 3.3 day period.
In addition a triple object containing an O8\,If star is present (Garcia \& Mermilliod
2001\nocite{2001A&A...368..122G}).
\smallskip
\newline
\noindent 
In conclusion, the young clusters located within 65\,pc of an 
O-type field star (Table\,\ref{runw2}) show the 
required characteristics (as far as we know) to be the possible hosts
from which a massive star can be dynamically ejected (e.g. Clarke \& Pringle 1992).

On the other hand, for the 15 runaway candidate stars discussed in Sect.\,\ref{space}
and \ref{zdist} it is much less likely that there exists a 
young cluster within the drift distance of 65\,pc. These candidate OB-runaway 
stars should have space velocities that would move 
them much further away from their birth sites. Indeed, this is 
borne out by our search that reveals no young cluster around any 
such object (see Paper\,I).

\section{Discussion}
\label{disc}
Whether the generally observed relation between massive stars and
clusters has a physical origin or is simply a statistical phenomenon
is here reconsidered using our analysis of the sample of 43
O-type stars in the Galactic field. Assuming that all high-mass stars
form in clusters, we have attempted to explain the current location of
these stars by exploring two possibilities: (1) they are in fact
members of small (or embedded) stellar clusters that have not been
detected previously; (2) they have presumably been ejected from a 
cluster birth site. These two explanations for the current isolated
location are found not to be satisfactory in a small number of cases. This
indicates that a small percentage of high-mass stars may indeed form in an 
isolated way (Li et al. 2003).

\subsection{The observed number of field O stars}
In paper\,I we have shown that 12\% (5/43) of  the massive
field stars are found located in stellar density enhancements. 
In two cases, these newly detected
clusters appear to be populous enough that the presence of a high-mass
star is consistent with what is expected from a standard IMF.  In
the other three cases, the detection is more marginal. In particular
the cluster marginally detected near HD\,57682 could be a false
detection, given the stars' high proper motion (see Sect\,3.1).

For the remaining stars we have sought an interpretation based on the
dynamical ejection from young stellar clusters, adopting less
stringent but still acceptable criteria in order to identify possible
runaway stars. Using Hipparcos proper motions, we have identified 7
field O stars with peculiar tangential motion normalized to one
component greater than $\rm 40$ km\,s$^{-1}$.  Eight additional stars
were found to exceed a distance of 250 pc from the Galactic
plane. Such a distance corresponds to the limit of the observed
vertical distribution of OB associations and to the maximum extent 
field O stars may wander given their average radial
velocity. Additionally, we have tentatively associated 7 field O stars
with clusters of age less than 10\,Myr, located within a projected
radius of 65\,pc. The resulting 22 field O stars with a possible
ejection history from a crowded environment are listed in
Table\,\ref{runw} and Table\,\ref{runw2}.  In order to identify the
genuine sample of isolated field O stars, we must discard HD\,135240, and
HD\,135591 that have been suggested to be part of the stellar group
Pis\,20 (Mel'Nik \& Efremov 1995) and the object HD\,113658 that is
reported to be member of the Cen\,OB1 association in the Luminous Star Catalogue
(Humphreys \& McElroy 1984\nocite{1984ApJ...284..565H}; see Paper\,I).

The presented statistical analysis reduces the number of field O stars to 11
objects whose current isolated location may correspond to their actual birth
location. They are given in Table\,\ref{single}. We note that all stars except
two are in the Hipparcos dataset, and were therefore not found to have high
proper motions in Sect\,3.1. The absolute number of 11 O-type field stars
corresponds to 6\% of the complete Galactic O-type star population with
$V<8^{m}$, considering its total number of 193 O-type stars (M98).  The
percentage can be refined by considering the fact that if indeed the field stars
in Table\,\ref{single} have formed as single objects, traces of star formation
(dark clouds, emission nebulae, etc.) in the surrounding field may still be
present. Our search in Paper\,I for such tracers resulted in positive detections
near four stars. They have been marked by an asterisk in Table\,\ref{single} and
would present the best examples for isolated Galactic high-mass star
formation. However high-mass stars can eject enough momentum (mechanical and
radiative) in the surrounding medium to efficiently remove hints of their
formation in $10^{6}$ years. Therefore the absence of star formation tracers
near the remaining 7 stars does not exclude that the field O star has not formed
in situ. We conclude that a refined percentage of $4\pm2\%$ of the high-mass
stars in the Galaxy may have formed in isolation. The current
incompleteness in the census of the embedded massive star population will not
change this ratio; both the embedded massive stars in clusters and the embedded
isolated massive stars will likely suffer from similar incompleteness
percentages. We see thus that 
nearly 95\% of the Galactic O star population is located in a cluster or OB
association or can be kinematically linked with clusters/associations. We
conclude that in the solar neighborhood, the probability of the formation of a
high-mass star in special conditions seems unlikely, but cannot be completely
excluded.

\begin{table}
 {
  \begin{center}
 \caption[]{The final 11 O-type field stars that could not be associated with clusters. 
An asterisk indicates if the surrounding field shows signs of star formation 
as reported in Paper\,I. The abbreviations are the same as in Table\,1 and 2.}
  \begin{tabular}{lcc}
\hline
\hline
Name & V.M. & S.M.\\
\hline
HD\,39680        & OPT    & C    \\
HD\,48279$^{*} $ & OPT    & C    \\
HD\,96917        & single & SB1: \\
HD\,112244       & VB     & SB1: \\
HD\,120678       & single & U    \\
HD\,123056$^{*}$ & single & C    \\
HD\,124314$^{*}$ & VB     & SB1: \\
HD\,154811       & single & C    \\
HD\,165319$^{*}$ & single & C    \\
HD\,193793       & VB     & SB2O \\
HD\,202124       & single & C    \\
\hline
 \end{tabular}
\label{single}
\end{center}
}
\end{table}

\subsection{Comparison with the predicted number of field O-type stars}
In this subsection we attempt to calculate the expected number of isolated 
field O-type stars, based on the assumption that stars form in clusters with  
a richness that is distributed according to a continuous power law down to 
the smallest clusters containing only one member.

That the number of stellar clusters are distributed in mass according to a power
law $\rm d{\it N}/dM_{cl}\sim M_{cl}^{- \beta_{CMF}}$ is becoming an established
idea. The cluster mass function (CMF) in different astrophysical environments
like super star clusters (Zhang \& Fall 1999)\nocite{1999ApJ...527L..81Z} or
globular clusters (Harris \& Pudritz 1994)\nocite{1994ApJ...429..177H} are
consistent with a value of $\rm \beta_{CMF}=2$ (see also Elmegreen \& Efremov
1997\nocite{1997ApJ...480..235E}). Recently Oey, King \& Parker
(2004)\nocite{2004AJ....127.1632O} showed that in the Small Magellanic Cloud
(SMC) the number of clusters/associations also appears to be distributed like a
power law with the richness in OB members ($N_{*}$), i.e. $dN/dN_{*}\sim
N_{*}^{- \beta}$. Empirically they found an index for $\beta= 2$ similar to the
distribution of the number of clusters by mass.

\begin{figure*}[t]
 \includegraphics[height=8cm,width=8cm]{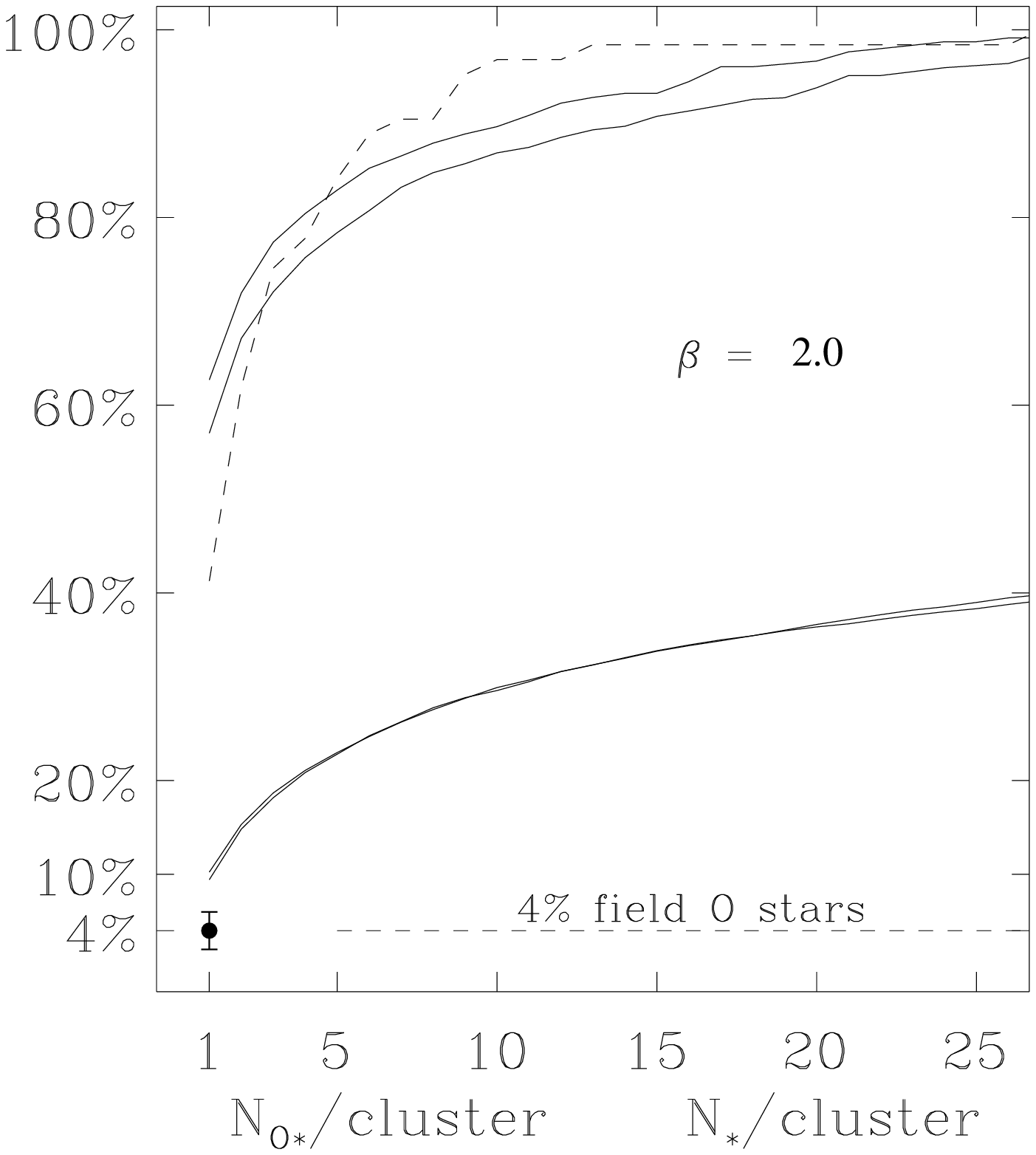} 
 \includegraphics[height=8cm,width=8cm]{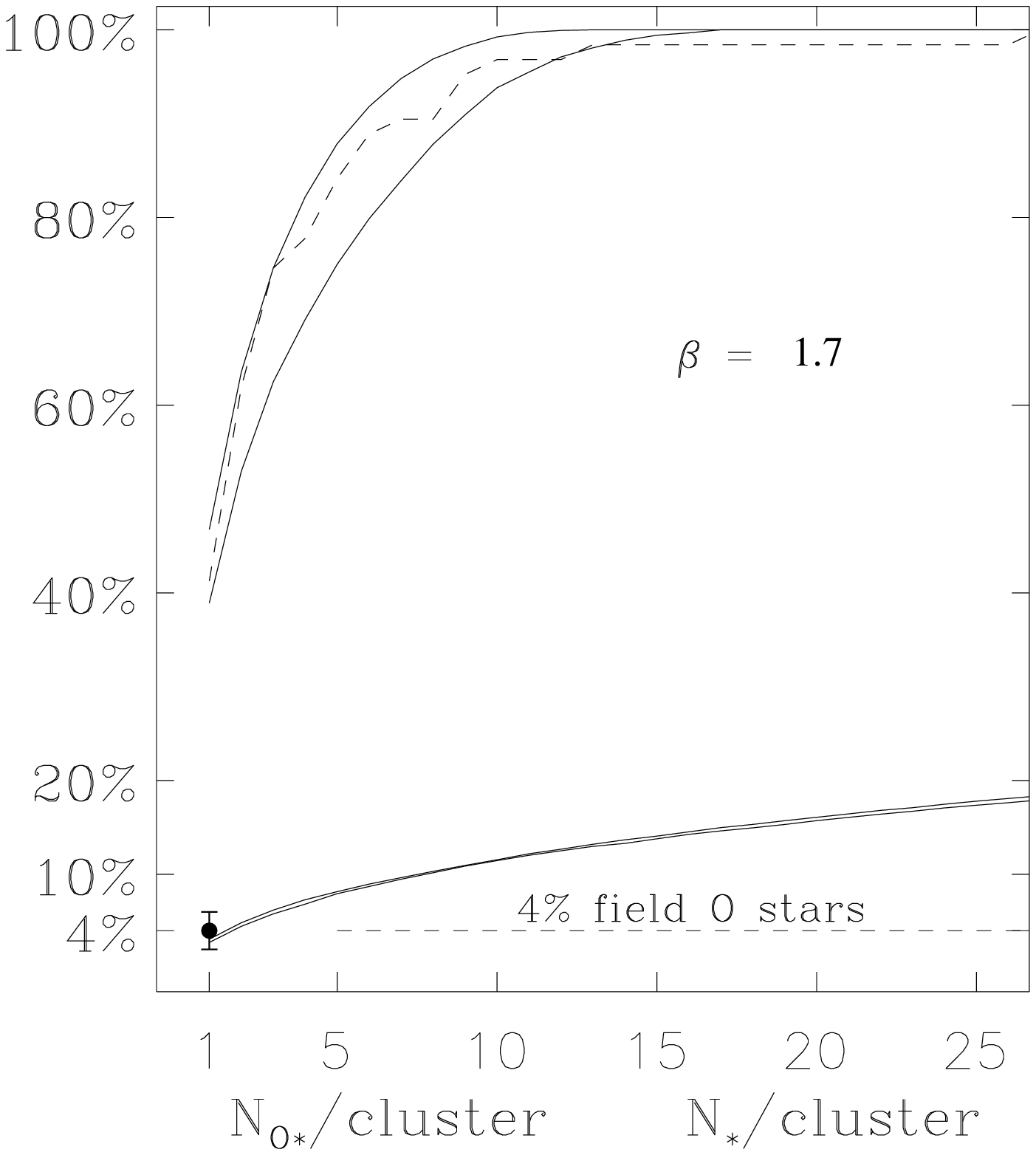}
 \caption[]{Both panels show the calculated cumulative distribution for the total number of {\it stars per cluster}
 (lower full curves) and the number of {\it O-type stars per cluster} (upper
 full curves). The simulations have been done for 2 stellar IMFs (see text),
 therefore there are 2 upper and 2 lower full curves, although the latter are
 hardly distinguishable. An O-type star in the simulation is a star more
 massive than $\rm 17.5\,M_{\odot}$. The dashed curves correspond to observed number
 fraction for Galactic OB-associations.  The point with the errorbar corresponds
 to the observed number of Galactic field O-type stars, following the analysis
 in this paper. The left panel is for a CMF with slope $\beta= 2.0$ and the right panel for
 $\beta= 1.7$. The right panel fits the observations.}
 \label{cum} 
\end{figure*}

An important finding by Oey et al.  is that the cluster distribution by $N_{*}$
can account for the cluster richness distribution for all observed OB
clusters/associations in the SMC down to clusters containing only one OB
star. One can extrapolate the Oey et al. result by assuming that (1) high mass
and low mass stars are all formed in clusters, and (2) the cluster richness
distribution follows a single power law, down to ``clusters'' containing a
single star. This assumption allows one to calculate the expected number of
isolated field O stars, by using a Monte Carlo simulation that draws cluster
richness from a certain power law distribution and populate each cluster
adopting a stellar IMF.  We note that the cluster distribution function we adopt
is by number of stars $N_{*}$ contained in each cluster and not by the total
mass of the cluster and these two distributions are in principle
different. However in the following we continue to refer to this distribution as
CMF.

In Fig.\,\ref{cum} the results of the Monte Carlo simulations are
presented in the form of cumulative distributions. The distributions
in each panel correspond to runs involving $0.5\times10^{6}$
clusters. Both panels show the cumulative distribution of two
different parameters as indicated on the horizontal axis. The first
parameter is the number of O-type stars per cluster (upper full curves
in both panels). The second parameter is the stellar richness for each
O-type star cluster, i.e. a cluster where the most massive member is
of O-type\footnote{An O-type star in this case would be a star more
massive than $\rm 17.5\,M_{\odot}$.}  (lower full curves in both
panels). There are two upper and two lower curves  corresponding to two
different stellar IMFs, viz. Salpeter (1955)\nocite{1955ApJ...121..161S} and Kroupa
(2002)\nocite{2002Sci...295...82K}. The calculations are found to
depend critically on the mass of the most massive cluster; more
massive clusters have more O-type stars and therefore the cumulative
distribution will rise more slowly. Observed quantities are
represented in each panel by dashed lines. The cumulative distribution
of the number of O-type stars per Galactic OB-association (corresponding to the upper curves) 
is derived from the catalogue of M98, where we took a pure statistical
approach, in that a O-type binary logically counts for two objects.

The left panel in Fig.\,\ref{cum} shows the results for a CMF with a
slope of $\beta= 2$. We applied the value of the most massive OB-association
in the sample of M98 in the computation. The largest association in M98 is
\object{Sco\,OB\,1} that contains 27 O-type stars, equivalent to $\rm\sim1.2\times10^{4}\,M_{\odot}$. 
The upper full curves indicate that
in this case $\sim60\%$ of all massive clusters/OB-association contain
only a single O-type star. The observed distribution (dashed line)
indicates a somewhat lower percentage amounting to $\sim40\%$. We see therefore that
a CMF of $\beta= 2$, produces too many single O-type star
OB-associations. The same cluster size distribution also predicts that
about 10\% of the O-type stars are single stars (lower curves, left
panel). Comparing this number with the rederived statistics in the
previous subsection, i.e. $4\pm2\%$, we see that this distribution
predicts too many single isolated O-type stars. 

In the right panel we try to fit the observations and find a best fit for a CMF
with a slope of $\beta= 1.7$. The fit requires a maximum number of stars per
cluster to be $3\times10^{3}$.  As far as the distribution of the number of
O-type stars per OB-association {\it and} the number of isolated O-type stars is
concerned, we conclude therefore that the Galactic data for O-type stars is
better fitted by a CMF with a slope of $\beta= 1.7$. This result differs from
the value of $\beta= 2$ for the distribution of the number of OB-type stars per
association in the SMC (Oey et al. 2004). On the other hand, a power law
function of $\beta= 1.7$ is also preferred by Bonnell \& Clarke
1999\nocite{1999MNRAS.309..461B} in their calculations to fit the empirical
richness distribution of the small-scale clusters near HAeBe stars (Testi et
al. 1999\nocite{1999A&A...342..515T}), indicating possibly a difference in the
the distribution of cluster richness between SMC and Galaxy.

\subsection{The field IMF of massive stars}
\label{youth}
If all stars were formed in stellar clusters
that have a continuous size distribution down to the smallest
clusters, then the production of isolated high-mass stars is primarily
dependent on the cluster mass function and to a much lesser extent
sensitive to stellar mass functions, as shown in the previous
subsection and clearly visible from the lower curve in
Fig.\,\ref{cum}.  The above assumption also implies that an IMF
derived from a field population will be found steeper than that of
individual clusters. The latter observation is borne out in two
recent papers by Kroupa (2004)\nocite{2004NewAR..48...47K} and Kroupa
\& Weidner (2003)\nocite{2003ApJ...598.1076K} discussing the IMF of
the Galactic field. They conclude that the IMF of the early type
field population is steeper than the IMF of a stellar cluster, with
$\rm \alpha_{field}\gta2.8$. As is correctly pointed out by the authors,
this does not necessarily constitute a different star formation mode,
but could be considered as the convolution of the cluster size and stellar
mass distribution functions.
 
Similar reasoning may hold for the very steep mass functions found for the
general stellar field in some external galaxies. These steep mass functions have
led to the suggestion of star formation modes in these galaxies different from
those observed in the Milky Way Galaxy. For example a very steep initial mass
function for the field of the Large Magellanic Cloud is derived by Massey et
al. (1995)\nocite{1995ApJ...438..188M} and Massey
(2002)\nocite{2002ApJS..141...81M}, based on their finding of high-mass field
stars. Also the bulge of M\,51 has been suggested to be capable of forming very
high-mass isolated stars, possibly the result of a different star formation mode
(Lamers et al.2002)\nocite{2002ApJ...566..818L}. The interesting question now is
whether the Galactic field O-type stars in Table\,\ref{single} have a formation
history similar to these extra-galactic objects, and whether the existence of
these objects could again be the combined effect of a stellar mass function and
a cluster size function. Proper motion observations for this particular set of
stars are especially warranted.

\section{Conclusion}
\label{concl}
Starting from the idea to test whether all massive stars form in
clusters, we aimed at elucidating the formation mechanism of Galactic
O-type field stars. Our two step approach consisted in a search for
as yet undetected stellar clusters hosting the seemingly
isolated high-mass stars (Paper\,I), and by
testing whether the field O-type stars have undergone a dynamical
ejection event from a young stellar cluster (this paper). The conclusion drawn in
Paper\,I is that the field O-type stars are for the large majority
isolated; the search for clusters resulted in the detection of only 5
possible clusters near a total number of 43 O-type field stars. In
this paper we conclude that a high fraction of the O-type field stars
(22/43) can be considered to be runaway star candidates, based on
their present or former peculiar space velocity, and the vicinity
of some of them to very young clusters ($\rm <10\,Myr$). 

The main conclusion is that at present $4\pm2\%$ of the complete population of
Galactic O-type stars cannot be traced to a formation in a cluster/OB-association. These stars
could be the Galactic analogues of isolated high-mass stars found in the general
stellar field of the Magellanic Clouds (e.g. Massey 2002).  If true, this result
implies that there is a small percentage of high-mass stars that may form in
isolation. In fact statistical arguments using the IMF and a cluster mass
function can actually reproduce such a percentage assuming that all stars are
formed in  clusters, that follow a universal cluster distribution (by $N_{*}$)
with a slope of $\beta= 1.7$ down to clusters with a single member. Statistically the
presence of high-mass stars in the field may represent the combined effect of a
universal cluster size distribution {\it and} a stellar mass function. In
physical terms, our study supports the conclusion about the rarity of the
formation of a high-mass star outside a stellar cluster, but this mode
(i.e. outside clusters) of high-mass star formation in the Galaxy cannot be excluded by our 
analysis presented here.

\begin{acknowledgements}
WJDW acknowledges financial assistance from the European Union Research
Training Network `The Formation and Evolution of Young Stellar Clusters'
(RTN1-1999-00436).  This research has made use of the Sim\-bad database,
operated at the Centre de Don\-nees Astro\-no\-miques de Stras\-bourg (CDS), and
of NASA's Astrophysics Data System Biblio\-graphic Services (ADS).
\end{acknowledgements}


\appendix
\section{Clusters near HD\,52533 and HD\,195592}
\label{appb}

HD\,52533 is surrounded by a cluster with an estimated density of $\rm
log\,(\rho_{I_C})=2.5\pm0.9$ stars per cubic parsec that is complete to
a minimum mass of $\rm \sim 0.8M_{\odot}$. This estimate is based on the $\rm I_C$
values converted to a volume density using the observed radius.  The
cluster stellar density is comparable to that of the richest clusters near
early-type HBe stars like MWC\,297 (Testi et al. 1999).

HD\,195592 is a particularly interesting case as the associated IRAS source
has a ``bubble'' or ``bowshock'' shape (Van Buren
1995\nocite{1995AJ....110.2914V}), typically found in runaway objects. The
fact that HD\,195592 is of luminosity class Ia does not affect the
interpretation that in an evolutionary sense it is closer to its origin than
to its dimise. For example, if we examine the MK spectral types of the Ori
OB1 association, for a single spectral type (O9.5) we find all luminosity
classes present as members (G87). Considering HD\,195592, we also note that
the density map (see Fig.\,16 of Paper\,I) shows a second cluster centered on
the bright early-type star to the NE of HD\,195592.  Unlike HD\,52533,
HD\,195592 has a relatively high extinction ($A_{\rm V}\sim3^{m}$), easily
seen in optical images that reveal the ionized emission originating from the
swirls of interstellar gas in the region. Since a single O5 star can
completely disperse a $\rm 10^4M_{\odot}$ molecular cloud in about 1 Myr
(Yorke 1986\nocite{1986ARA&A..24...49Y}), such a cloud could well be the
remnant of the birth environment. All this indirect evidence suggests a
relatively young age for HD\,195592.

\end{document}